\documentclass{aa}
\usepackage{graphics}
\input psfig.sty

\begin{document}


\title{The HST survey of the B2 sample 
of radio-galaxies: optical nuclei and the FR~I/BL Lac unified scheme
\thanks{Based on observations with the NASA/ESA Hubble Space Telescope, 
obtained at the
Space Telescope Science Institute, which is operated by AURA, Inc.,
under NASA contract NAS 5-26555 and by STScI grant GO-3594.01-91A}}

\author{
A. Capetti \inst{1} \and 
A. Celotti \inst{2} \and
M. Chiaberge\inst{3} \and 
H.R. de Ruiter\inst{4,5} \and 
R. Fanti\inst{5,6} \and 
R. Morganti\inst{7} \and 
P. Parma\inst{5}
}

\institute{Osservatorio Astronomico di Torino, Strada Osservatorio 20, 
I-10025 Pino Torinese, Italy 
\and
SISSA/ISAS, Via Beirut 2-4, I-34014 Trieste, Italy
\and
Space Telescope Science Institute, 3700 San Martin Drive, 21218 Baltimore MD,
U.S.A.
\and
Osservatorio Astronomico di Bologna, Via Ranzani, 1, I-40127 Bologna, Italy
\and
Istituto di Radioastronomia, Via Gobetti 101, I-40129 Bologna, Italy
\and
Dipartimento di Fisica dell'Universit{\`a} di Bologna, Via Irnerio 46, 
I-40126 Bologna, Italy
\and
Netherlands Foundation for Research in Astronomy, Postbus 2,
NL-7990 AA, Dwingeloo, The Netherlands}

\offprints{capetti@to.astro.it}

\date{Received ...; accepted ...}

\titlerunning{Optical nuclei in B2 radio-galaxies}
\authorrunning{Capetti et al.}  

\abstract{We examine the optical properties of the nuclei of low luminosity
radio-galaxies using snapshot HST images of the B2 sample.  In
agreement with the results obtained from the analysis of the brighter
3C/FR~I sample, we find a correlation between fluxes (and
luminosities) of the optical and radio cores.  This provides further
support for the interpretation that the optical nuclear emission in
FR~I is dominated by synchrotron emission and that accretion in these
sources takes place in a low efficiency radiative regime.  In the
framework of the FR~I/BL Lacs unified scheme,
we find that the luminosity difference between FR~I and BL Lac nuclei
can be reproduced with a common beaming factor in both the radio
and the optical band, independent of the extended radio
luminosity, thus supporting such a scenario.
The corresponding bulk Lorentz factor is significantly smaller than 
is expected from observational and theoretical considerations in BL
Lacs: this can be interpreted as due to
a velocity structure in the jet, with a fast spine surrounded by a
slower layer. 
}

\maketitle

\section{Introduction}
\label{intro}

The presence of a radio-source represents a common manifestation of
nuclear activity associated to elliptical galaxies, in particular for
the brightest members of this class.  For example, among galaxies
brighter than $M_{\rm B} < -21$, more than 20 \% have radio
luminosities $L_{\rm 408 MHz} > 10^{23.5}$ W Hz$^{-1}$ (Colla et
al. \cite{colla}). Due to the steepness of the radio luminosity
function, most of them are low luminosity radio-sources and show the
characteristic edge-darkened FR~I radio morphology (Fanaroff and Riley
\cite{fr}).  However, in observing bands other than the radio,
emission related to their active nuclei remains largely
elusive. Optical spectra of low luminosity radio-galaxies are in fact
dominated by the stellar component of the host galaxy with only faint
narrow emission lines, while permitted broad lines are seldom
detected. Similarly, most of their X-ray emission originates in the
ambient thermal plasma. With the limited information available, we
cannot effectively constrain the physical properties and the emission
processes at work in these AGNs, and  to determine how these
sources fit into the AGN unification schemes: one has to rely on the
properties of the host galaxies, of the environment or of the
extended radio structure.

Significant progress in our understanding of low luminosity
radio-galaxies has been achieved recently thanks to  high
resolution HST and Chandra 
images, which enabled us to isolate their genuine
optical and X-ray nuclear emission. In HST images unresolved nuclear sources
are detected in the great majority of the 33 FR~I belonging to the 3CR
sample (Chiaberge et al. \cite{chiaberge99}, hereafter
CCC99).  The optical flux density of these Central Compact Cores (CCC) shows a
striking linear correlation with the radio core, arguing for a
common non--thermal synchrotron origin.  In five sources, for which it
is possible to estimate the viewing angle based on the inclination of
their circumnuclear dusty disks, the luminosity of the central source
shows a suggestive dependence on the radio galaxy orientation, as
qualitatively expected if the optical emission is indeed produced in a
relativistic jet (Capetti and Celotti \cite{ac2}).  The high rate of
CCC detection suggests that a geometrically thick torus can be present
at most in a minority of low luminosity radio galaxies.  CCC fluxes
also represent upper limits to any thermal/disk emission that
translate (for a $10^9 M_{\sun}$ black hole) into a fraction $\sim
10^{-7}-10^{-5}$ of the Eddington luminosity, suggesting that
accretion takes place at low rate or in a low efficiency radiative
regime.

This information also offers a new possibility of testing the FR~I/BL
Lac unification scheme, by directly comparing the optical nuclear
properties of radio galaxies with their putative aligned (beamed)
counterparts, analogously to the procedure followed in the radio band
(Kollgaard et al.  \cite{koll96}). From this comparison Chiaberge et
al. (\cite{chiaberge00b}, hereafter CCCG00) 
found that the difference in luminosity (in
radio and optical bands) between FR~I nuclei and BL Lac is
significantly smaller than  would be expected, in the frame of a
simple one--zone model, from the high bulk Lorentz factor of BL Lacs
jets implied by observational and theoretical considerations
(e.g. Dondi \& Ghisellini \cite{dondi}, Ghisellini et al. \cite{gg98},
Tavecchio et al. \cite{taold}).  In order to
reconcile these results with the unification scheme, they suggested
that a velocity structure is present in the jet, with a fast spine
surrounded by a slow layer, as already suggested by other evidence on
larger scales (e.g. Laing 1993, Laing et al. \cite{laing99}).

We recently obtained HST images for more than half of the B2 sample of
low luminosity radio galaxies (see Capetti et al. \cite{capetti00},
hereafter Paper I). Therefore, it is now possible to complement the
analysis performed for the 3C sources with a study of sources at
lower radio luminosities.  This extension will allow us to test the
general validity of these results for the radio galaxy population
and to explore in more detail the relationship between FR~I and BL
Lacs.
 
Thus in this paper we focus on the properties of their optical nuclei.
In Sect. \ref{thesample} we briefly present the properties of the B2
sample and the HST observations on which this study is based; the
optical nuclear properties of the galaxies are described in
Sect. \ref{CCC} where we also quantify the contribution of their
central optical sources.  In Sect. \ref{discussion} we discuss the
implications of our results, which are summarized in
Sect. \ref{summary}. For consistency with Paper I we use H$_{\rm o}$=
100 km s$^{-1}$ Mpc$^{-1}$ and $q_0=0.5$.

\section{The sample and the HST observations}
\label{thesample}

\begin{table*} 
\caption{Nuclear properties of the sample.}\label{tab1}
\hspace{1.5cm} 
\scriptsize
\begin{flushleft}
\begin{tabular}{l l l c c r c c} \hline\noalign{\smallskip}
Name & \multicolumn{2}{c}{Alternative ID} & Redshift & $\log P_t$ & $\log P_c$ & Optical & Radio Morph. \\
     &     Radio  &     Optical       &          &   W/Hz     &  W/Hz      & erg/s cm$^{-2} \AA^{-1}$ &  \\
\noalign{\smallskip}
\hline\noalign{\smallskip}
0034+25 &         & UGC00367 & 0.0321 & 23.20 &    21.62 & Dusty            &  FR~I \\ 
0055+26 & 4C26.03 & NGC0326  & 0.0472 & 24.61 &    22.30 & $<$ 1.0e-18      &  FR~I \\
0055+30 &         & NGC0315  & 0.0167 & 24.08 &    23.24 & 3.1e-17$^{d}$    &  FR~I \\
0104+32 & 3C031   & NGC0383  & 0.0169 & 24.21 &    22.45 & 1.5e-17$^{a,d}$  &  FR~I \\
0116+31 & 4C31.04 &          & 0.0592 & 24.95 &    22.34 & Dusty            &  core \\
0120+33 &         & NGC0507  & 0.0164 & 22.30 &    20.60 & $<$ 0.6e-18      &  FR~I \\
0149+35 &         & NGC0708  & 0.0160 & 22.33 &    21.13 & Dusty            &  FR~I \\
0648+27 &         &          & 0.0409 & 23.62 &    23.00 & Dusty            &  core \\
0708+32 &         &          & 0.0672 & 23.51 &    22.85 & $<$ 1.5e-18      & - \\
0722+30 &         &          & 0.0191 & 22.67 &    22.02 & Spiral           & disk     \\
0755+37 & 3C189   & NGC2484  & 0.0413 & 24.49 &    23.59 & 3.2e-17          &  FR~I \\
0908+37 &         &          & 0.1040 & 24.84 &    23.46 & 3.7e-18$^{d}$    &  FR~I/II \\
0915+32 &         &          & 0.0620 & 24.00 &    22.50 & Dusty            &  FR~I \\ 
0924+30 &         &          & 0.0266 & 23.52 & $<$20.50 & $<$ 0.5e-18      &  FR~I \\
1003+26 &         &          & 0.1165 & 24.01 &    22.10 & $<$ 0.2e-18      & -  \\
1003+35 & 3C236   &          & 0.0989 & 25.78 &    24.00 & $<$ 3.5e-18      &  FR~II \\
1005+28 &         &          & 0.1476 & 24.25 &    22.60 & $<$ 0.6e-18      &  FR~I/II \\
1101+38 &         & MRK 421  & 0.0300 & 23.97 &    23.32 & 1.15e-14      &  core \\
1113+24 &         &          & 0.1021 & 23.65 & $<$22.50 & $<$ 0.2e-18      &  FR~I/II \\
1204+34 &         &          & 0.0788 & 24.47 &    22.88 & $<$ 3.7e-18      &  FR~II \\
1217+29 &         & NGC4278  & 0.0021 & 21.24 &    20.48 & 2.4e-17          &  core \\
1251+27 & 3C277.3 & Coma A   & 0.0857 & 25.37 &    22.95 & 1.5e-18$^a$      &  FR~II \\
1256+28 &         & NGC4869  & 0.0224 & 23.05 &    21.08 & Dusty            &  FR~I \\ 
1257+28 &         & NGC4874  & 0.0239 & 23.08 &    20.82 & $<$ 0.2e-18      &  FR~I \\
1321+31 &         & NGC5127  & 0.0161 & 23.85 &    21.77 & Dusty            &  FR~I \\
1322+36 & 4C36.24 & NGC5141  & 0.0175 & 24.55 &    22.21 & Dusty            &  FR~I \\
1339+26 & 4C26.41 & UGC08669 & 0.0757 & 24.30 & $<$22.40 & Dusty            &  FR~I \\
1346+26 & 4C26.42 &          & 0.0633 & 24.55 &    23.37 & 3.4e-18          &  FR~I \\
1347+28 &         &          & 0.0724 & 24.05 &    22.34 & Dusty            &  FR~I/II \\
1350+31 & 3C293   & UGC08782 & 0.0452 & 25.03 &    23.34 & Dusty            &  FR~I/II \\
1357+28 &         &          & 0.0629 & 24.03 &    22.45 & Dusty            &  FR~I \\ 
1422+26 &         &          & 0.0370 & 24.00 &    22.46 & $<$ 6.9e-18      &  FR~I \\
1430+25 & 4C25.46 &          & 0.0813 & 24.20 & $<$21.90 & $<$ 2.2e-18      &  FR~I \\
1447+27 &         &          & 0.0306 & 22.78 &    22.76 & $<$ 4.1e-18      &  core \\
1450+28 &         &          & 0.1265 & 24.50 &    23.01 & $<$ 1.3e-18      &  FR~I \\
1457+29 &         &          & 0.1411 & 25.22 & $<$23.40 & Dusty            &  FR~I \\
1455+28 & 4C28.38 &          & 0.1470 & 24.89 & $<$23.00 & Dusty            &  FR~II \\
1502+26 & 3C310   &          & 0.0540 & 25.36 &    23.40 & 3.5e-18$^a$      &  FR~I \\
1511+26 & 3C315   &          & 0.1078 & 25.34 & $<$24.24 & Complex          &  FR~I \\  
1512+30 &         &          & 0.0931 & 23.82 &    21.60 & $<$ 0.4e-18      & -  \\
1521+28 & 4C28.39 &          & 0.0825 & 24.58 &    23.57 & 8.4e-18          &  FR~I \\
1525+29 &         & UGC09861 & 0.0653 & 23.98 &    22.04 & Dusty            &  FR~I \\
1527+30 &         &          & 0.1143 & 24.05 &    22.70 & $<$ 2.1e-18      &  FR~I \\  
1553+24 &         &          & 0.0426 & 23.57 &    23.02 & 1.4e-17          &  FR~I \\
1557+26 &         &          & 0.0442 & 22.81 & $<$22.80 & $<$ 9.0e-18      &  core \\
1610+29 &         & NGC6086  & 0.0313 & 22.93 & $<$21.02 & $<$ 0.9e-18      &  FR~I \\
1613+27 &         &          & 0.0647 & 24.03 &    22.64 & $<$ 1.5e-18      &  FR~I \\
1615+32 & 3C332   &          & 0.1520 & 25.79 &    23.43 & 8.8e-17$^a$   &  FR~II \\
1626+39 & 3C338   & NGC6166  & 0.0303 & 24.49 &    23.00 & 1.0e-17$^a$      &  FR~I \\
1658+30A& 4C30.31 &          & 0.0351 & 23.88 &    22.89 & $<$ 1.1e-17      &  FR~I/II \\
1726+31 & 3C357   &          & 0.1670 & 25.89 &    23.34 & $<$ 0.9e-18      &  FR~II \\
1827+32 &         &          & 0.0659 & 24.07 &    22.95 & $<$ 2.0e-18      &  FR~I \\
1833+32 & 3C382   &          & 0.0578 & 25.07 &    23.85 & 4.8e-15$^a$      &  FR~II \\
2116+26 &         & NGC7052  & 0.0164 & 22.79 &    22.08 & 2.2e-18$^{d}$    &  FR~I \\
2229+39 & 3C449   & UGC12064 & 0.0181 & 24.03 &    22.11 & 1.8e-17$^{a,d}$  &  FR~I \\
2236+35 &         & UGC12127 & 0.0277 & 23.47 &    21.74 & 1.7e-18          &  FR~I \\
2335+26 & 3C465   & NGC7720  & 0.0301 & 24.88 &    23.38 & 1.9e-17$^a$      &  FR~I \\
\noalign{\smallskip}
\hline
\end{tabular}
\end{flushleft}
$^a$ flux in the F702W (R) filter; $^d$ Central source in dusty galaxy.
\end{table*}

The B2 sample consists of 100 low luminosity radio-galaxies.  It is
constituted by bright galaxies (down to a limiting magnitude 
$m_{\rm v}$ = 16.5)
associated with radio sources from the B2 catalogue, which is complete
to 0.25 Jy at 408 MHz (Colla et al.~\cite{colla}; Fanti et
al.~\cite{fanti78}).  The sample spans the power range between
10$^{21}$ and 10$^{26}$ W Hz$^{-1}$ at 1.4 GHz with a pronounced peak
around 10$^{24}$ W Hz$^{-1}$ and therefore gives an excellent
representation of the radio source types encountered below and around
the break in the radio luminosity function 
(10$^{24.5 - 25}$ W Hz$^{-1}$ at 408 MHz Colla et al. 1975).  
Since it was selected at
low radio frequency, the sample is largely unbiased for orientation.
The B2 sources span a range in redshift between 0.002 and 0.176.

HST images are available for 57 B2 radio galaxies. In the course of our
HST snapshot program observations were obtained for 41 sources
(see Paper I for
details on the observations and data reduction). Each
source was observed using two broad band filters, namely F555W and
F814W (which approximately match the standard V and I filters) with an
exposure time of 300 s.  Public archival images for 16 additional
objects were mostly obtained with the F702W (R) filter (as
part of the 3CR snapshot programs, De Koff et al.  \cite{dekoff},
Martel et al. \cite{martel}).

While there is no bias in the selection of the sources that are part of the
snapshot program (they were chosen randomly as far as their radio and
optical properties are concerned), a bias might have been introduced
by the inclusion of the 16 archive sources.  However, the comparison
of various parameters of the observed sub-sample with those of the
sources that were not observed by HST shows that, except for a
marginal (within 2$\sigma$) difference in redshift and radio power
\footnote{The median redshift and radio power of the two sub-samples
are $ z = 0.055^{+0.008}_{-0.011}$, log $P_{\rm t} = 24.05^{+0.07}_{-0.03}$
and $ z = 0.067^{+0.005}_{-0.007}$, log $P_{\rm t} = 24.22^{+0.06}_{-0.03}$,
respectively.}, the sub-sample observed with HST can be considered as
well representative of the whole B2 sample.

The data of the observed sources relevant for this paper are given in
Table ~\ref{tab1}: in column 1 the B2 name, in columns 2 and 3 their
alternative radio and optical identifications, in column 4 the
redshift, in columns 5 and 6 the total and core radio power at
1.4~GHz, in columns 7 the optical nuclear flux density in the I band and in
column 8 the morphological classification of the extended radio
emission.

\section{Detection and photometry of the Central Compact Cores} 
\label{CCC}

\begin{figure} 
\resizebox{\hsize}{!}{\includegraphics{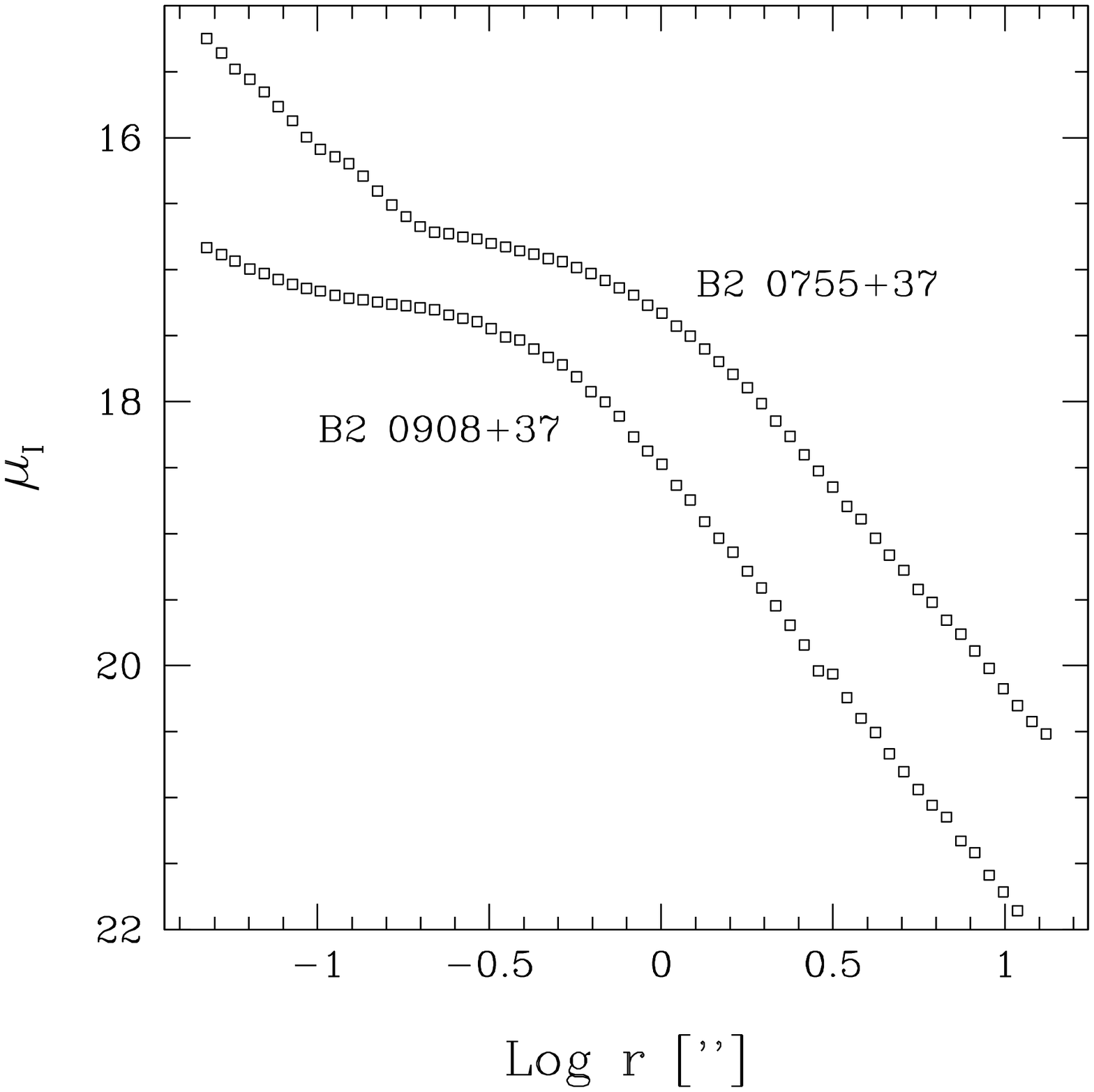}}
\caption{Radial brightness profile for B2 0755+37 and B2 0908+37.
Note the flat resolved galaxy core and the steepening of 
the profile caused by the nuclear source.}
\label{prof} 
\end{figure}

With the aim of determining whether, superposed to the galaxy
background, a CCC is present, we examined the nuclear brightness
profile of the B2 galaxies.  For the identification of optical cores
we adopted the same procedure described in CCC99, i.e. a source
brightness profile that, within 5 pixels from the center, shows a FWHM
consistent with the HST Point Spread Function ($\le 0\farcs08$). For
the bright elliptical galaxies harbouring the B2 sources this
procedure is particularly effective as they all show well resolved
central cusps characterized by shallow brightness gradients ($I(r)
\sim r^{-\gamma}$ with $\gamma \sim 0 - 0.3$).
Fig.~\ref{prof} presents profiles for two sources in which we detected
a nuclear source.

Clearly, this method can only be applied when the central regions of
the galaxy are not heavily affected by morphological peculiarities.
Nuclear dust features are seen in 20 B2 sources (de Ruiter et al. \cite{dust}) 
and they prevent a detailed analysis of the
innermost regions. Nonetheless in 5 of these
galaxies the presence of central unresolved components is readily
apparent from the visual inspection of the images and it is confirmed 
by the brightness profile behaviour; all of them are associated with
disky dust structures. Two additional sources are excluded
from the sample: B2~0722+30 which is a spiral galaxy and B2~1511+26
which presents a peculiar elongated morphology. 

From the brightness
profile analysis of the 35 remaining galaxies we found a CCC in 13 of them
(for a total of 18 central sources). 
In the other 22 cases we set an upper limit to the CCC component
defined, again following CCC99, as the light excess of the central 3x3
pixels with respect to the surrounding galaxy background.  We also
tested an alternative upper limit estimator, based on the subtraction
of a Point Spread Function opportunely scaled until it produces an
inversion in the galaxy profile, which yielded similar results.

Of the 18 unresolved optical sources, 2 are associated to B2~1615+32
(3C~332) and B2~1833+32 (3C~382), which are both Broad Line Radio
Galaxies with a clear FR~II radio morphology (Chiaberge et
al. \cite{chiaberge00a}, \cite{chiaberge01}); as expected, we also
detected a central source in B2~1101+38 (Mrk 421), a well known nearby
BL Lac object.  The remaining CCC are mostly associated to FR~I sources (12),
but we also found a CCC in 1 FR~II, 1 FR~I/II transition  source and 1 core source.
Six of them are in common with the 3C/FR~I sample (see Table 1).

We performed aperture photometry of the CCCs detected, adopting the
internal HST/WFPC2 flux calibration. The resulting measurements are
reported in Table ~\ref{tab1}, with errors ranging from 5\% to 25\%.
Images in two broad filters are available for most galaxies, but the
typical errors in the CCC photometry and the relatively small
wavelength range available do not allow  an
accurate estimate of the optical spectral indices in most cases
\footnote{The spectral index error scales as $\sigma_{\alpha}\sim 8.5
~ \sigma_{\rm F}/{\rm F}$, 
where $\sigma_{\rm F}/{\rm F}$ is the fractional error on the flux estimate,
and it amounts to typically $\sim$ 0.4 - 2}.

\section{Results and discussion} 
\label{discussion}

\begin{figure} 
\resizebox{\hsize}{!}{\includegraphics{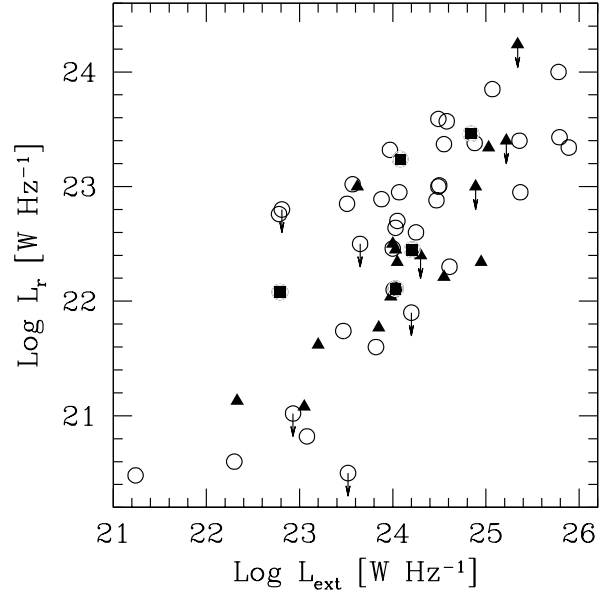}}
\caption{Core versus extended radio luminosities at 1.4 GHz of the B2
galaxies. Sources with obscured/complex optical morphology are marked as
filled triangle, while dusty galaxies with a central source are filled
squares; the remaining galaxies are plotted as circles.}
\label{dust} 
\end{figure}

Before discussing the properties of B2 cores, it is important to
establish the role of the sources with complex/obscured optical
morphology for which we could not estimate the nuclear optical
contribution, as their exclusion can in principle induce a significant
bias in the final sample.  This is particularly important for the
B2/HST sample as the incidence of nuclear dust appears to be
relatively high.  In Fig.~\ref{dust} we plot the radio core versus the
extended radio luminosities for all B2 studied, marking the location
of the complex/obscured galaxies with and without a central optical
source.  They are distributed over the whole range of extended
luminosities and they do not show that radio cores are significantly different
from the other galaxies with similar extended radio power.  We
quantitatively tested this result using the survival analysis package
{\it ASURV} (Isobe \& Feigelson \cite{asurv1}, \cite{asurv2}).  We
derived the correlations between $L_{\rm r}$ (core) and $L_{\rm ext}$ for the
three classes separately and no significant differences are found.
Thus, the exclusion of these sources will not affect significantly our
results and conclusions, at least as far as the radio properties are
concerned.

\subsection{The radio/optical plane for B2 radio galaxies cores} 
\label{correlation}

\begin{figure*}
\centerline{ \psfig{figure=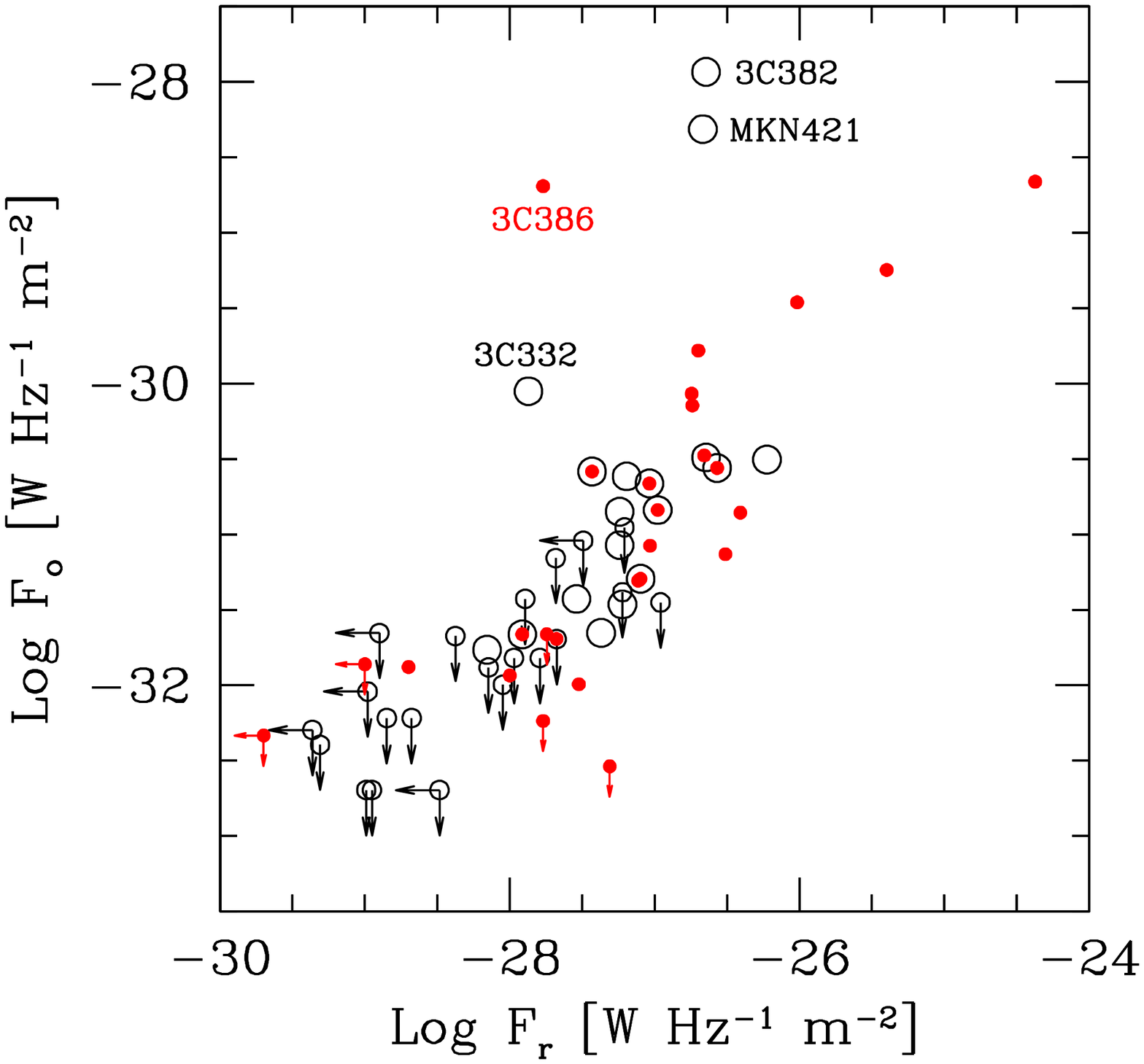,width=0.5\textwidth}
\psfig{figure=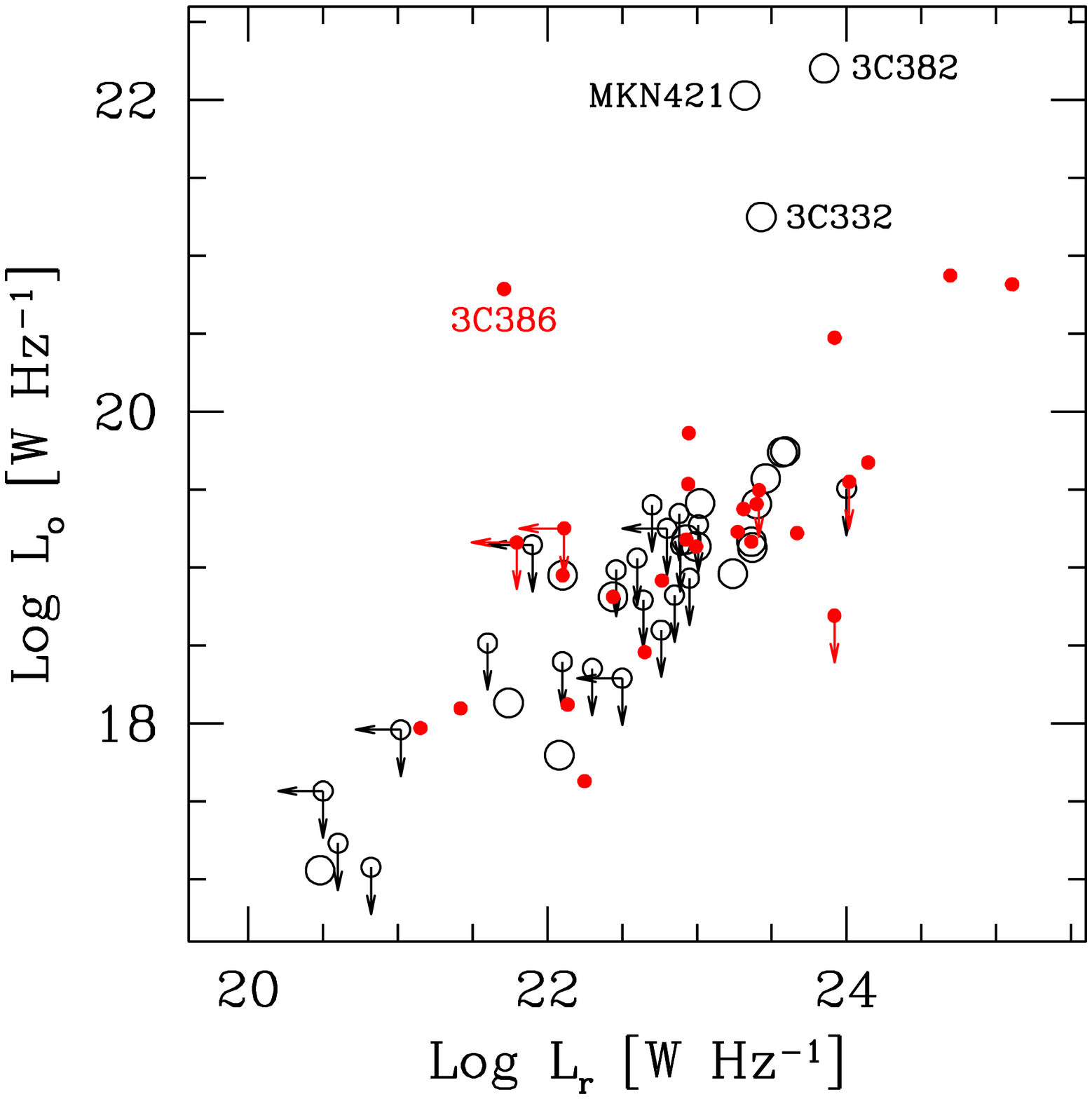,width=0.5\textwidth}}
\caption{Left panel: optical flux density of the CCC (at 8140 \AA),
versus radio core flux density (at 1.4 GHz).
B2 sources are represented by empty circles (large for detections, smaller
for upper limits), while the filled
circles are the 3C/FR~I sources from Chiaberge et
al. (\cite{chiaberge99}).  Right panel: same as the left panel for the
optical and radio luminosities.  The three sources with the highest
optical flux and luminosity are two BLRGs 
B2~1615+32 (3C~332) and B2~1833+32 (3C~382) and a BL Lac object 
B2~1101+38 (Mkn 421).}
\label{figcorr}
\end{figure*}

As discussed in the Introduction, the location of radio galaxies cores
in the radio/optical plane is a very useful tool to study the
properties of their nuclear emission.  In Fig. \ref{figcorr} we 
plot the radio versus optical fluxes and luminosities of the CCC for
the 40 B2 galaxies with CCC detection or upper limits together with
the data derived for the 3C/FR~I from CCC99.

CCC in B2 sources are located essentially in the same region of the
$F_{\rm r}$ - $F_{\rm o}$ and $L_{\rm r}$ - $L_{\rm o}$ 
planes as the 3C sources,and only
avoid the upper right corner corresponding to the highest fluxes
(and luminosities).  The only B2 sources which significantly deviate
are B2~1615+32, B2~1833+32 and B2~1101+38 which, as mentioned above,
are two BLRG FR~II and a BL Lac.  This agrees with the results of
Chiaberge et al. (\cite{chiaberge00a}, CCCG00) that both
BLRG and BL Lacs (in particular those belonging to the high energy
peak class
\footnote{Instead of classifying BL Lacs
according to their selection spectral band, we adopt the definitions
of high and low energy peaked BL Lacs (HBL and LBL, respectively),
which are based on the position of the (synchrotron) emission peak in
the spectrum and therefore more indicative of the physical
characteristics of the objects (Giommi \& Padovani \cite{giopad94},
Fossati et al.  \cite{gfos})}, as it is for B2~1101+38) present a 
large optical excess with respect to FR~I galaxies.  

Radio and optical emission of the B2 cores are clearly correlated 
and a survival analysis gives a probability that they are extracted from
a random population of only 0.0001. The best linear fit from ASURV
has a slope of
0.8 $\pm$ 0.1 consistent with the result derived by CCC99 for the 3C/FR~I
sample. It
confirms the findings of the analysis by CCC99 and
represents further evidence for a common synchrotron origin of the
radio and the optical emission in FR~I nuclei. 

\begin{figure*}
\centerline{
\psfig{figure=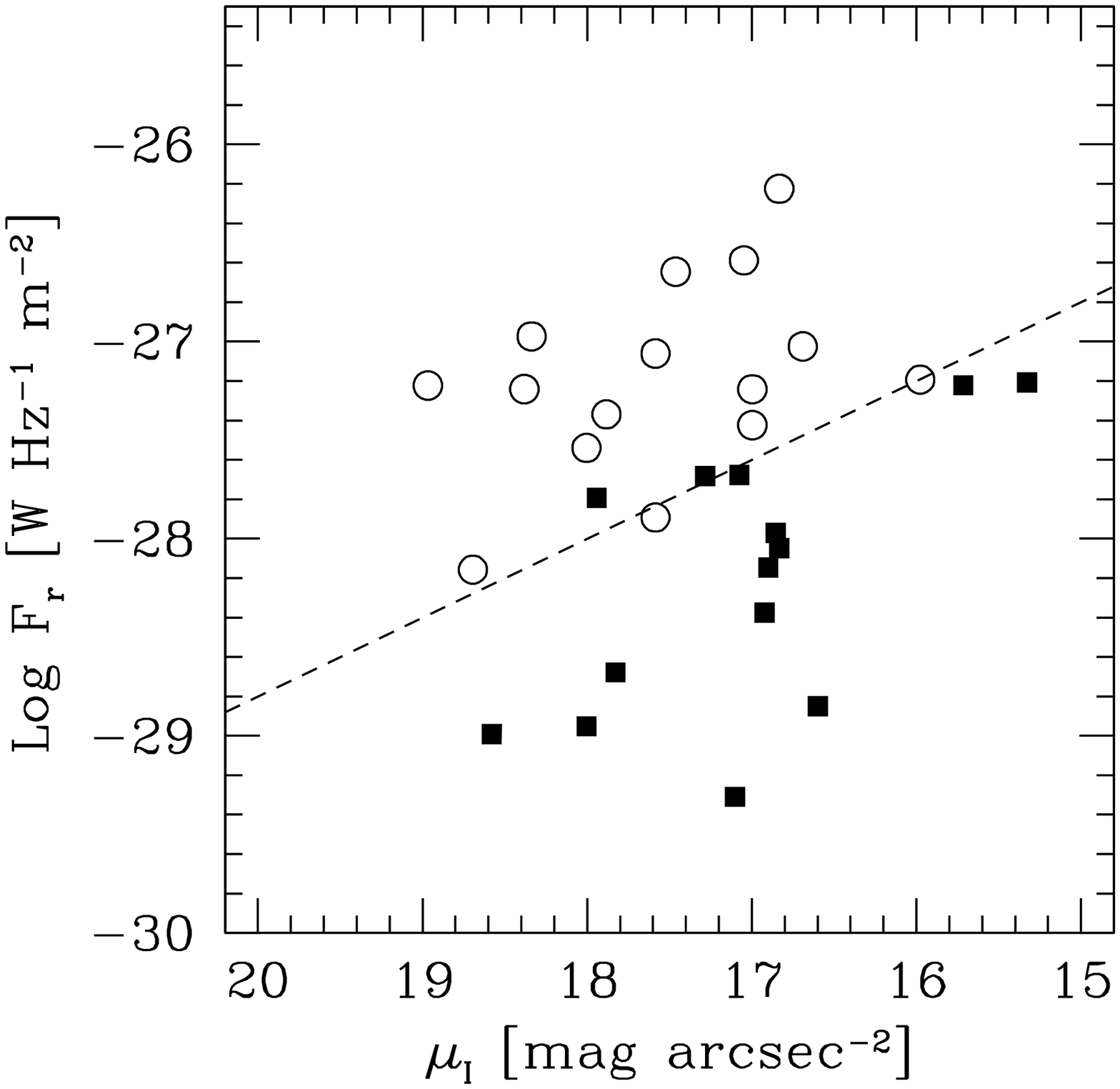,width=0.5\textwidth}
\psfig{figure=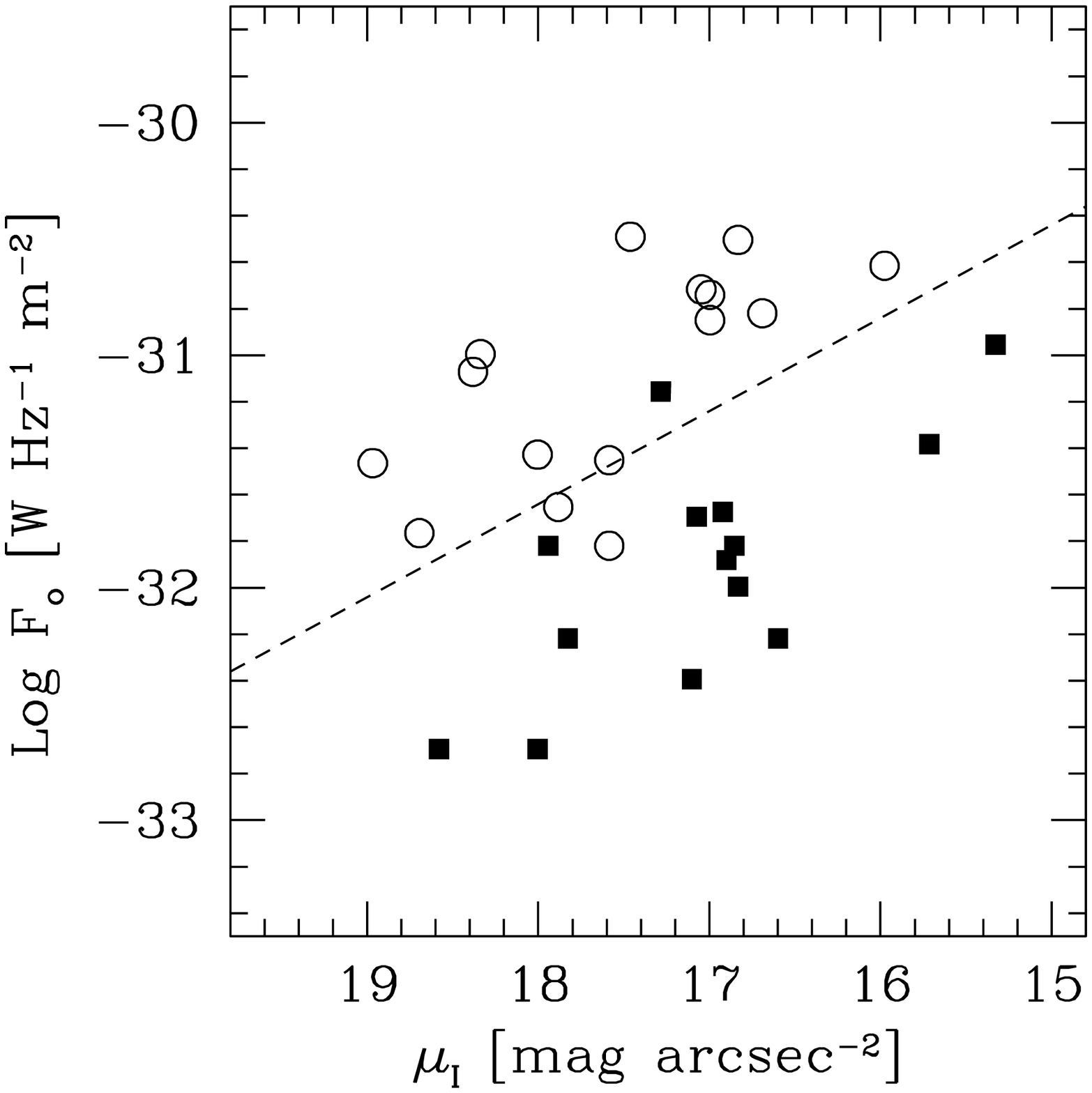,width=0.5\textwidth}}
\caption{Radio and optical core flux density 
versus central surface brightness of the host
galaxy. Objects with detected optical sources 
(excluding the 2 BLRG and the BL Lac object) are marked with open
circles, while squares represent galaxies with optical upper
limits. The dotted line marks a constant ratio between optical core
and galaxy emission (see text).}
\label{sb}
\end{figure*}

Galaxies for which we could only set upper limits to their optical
nuclear emission also appear to be consistent with this correlation,
but given their relatively high incidence, it is important to explore
in more detail their nature. In the following we test the idea that
their nuclei follow a similar radio/optical correlation as the
detected CCC, but their optical flux density is below our detection threshold.
In this case, our ability to detect a CCC will depend primarily on its
corresponding radio flux density, but also on its contrast with the host
galaxy. Indeed sources with and without detected CCC are well
separated in a plane in which we compare the radio core flux density with the
galaxy central surface brightness (see Fig. \ref{sb}). 
The separation occurs at an optical nuclear flux density
(derived assuming a radio-to-optical spectral index of 0.7, as
implied by the $L_{\rm o} -L_{\rm r}$ correlation)
that exceeds the galaxy emission within the central 0.1\arcsec x
0.1\arcsec. Undetected CCC are thus likely to be simply too faint to
be seen at the center of their host galaxies.  
This argues against alternative interpretations, such
as,  that they are obscured.

The inclusion of the B2 cores improves the coverage of the
radio/optical plane with respect to the 3C/FR~I sample alone, in
particular toward the lower luminosities.  This region is particular
important as here one might expect to see emerging the contribution of
radiation processes other than synchrotron emission from the jet.  The
presence of any additional component, unrelated to the radio emission,
should manifest itself as a flattening of the slope in the radio/optical
plane.  Clearly, this is not observed down to an optical luminosity of
$\sim 10^{18}$ W Hz$^{-1}$. For a 10$^9$ M\sun\ central black hole,
this corresponds to a fraction $\sim 3\times 10^{-8}$ of the Eddington
luminosity in the optical band. As already remarked in CCC99, this is
a clear evidence that accretion must occur at a very low rate and/or
radiative efficiency.

Of the FR~II and the transition FR~I/II sources (which are
represented by 7 and 5 galaxies, respectively) in addition to
the two BLRG already reported, only in two of them a CCC is detected
and this clearly prevents any comparison between the properties of
galaxies with different radio morphology.

\subsection{The FR~I / BL Lacs unification model} 
\label{extended}

Measurements of the optical nuclear emission in low luminosity radio galaxies
provided us with a new quantitative method of testing the FR~I/BL Lac
unification scheme.  The comparison between the beamed sources and
their putative counterparts must be restricted to objects of
properties which do not depend on orientation.  In particular, it is
important that they share a common range of extended radio luminosity
as this is plausibly linked to the energy carried by the relativistic
jet, which is at the origin of the nuclear emission.

\begin{figure*}
\centerline{
\psfig{figure=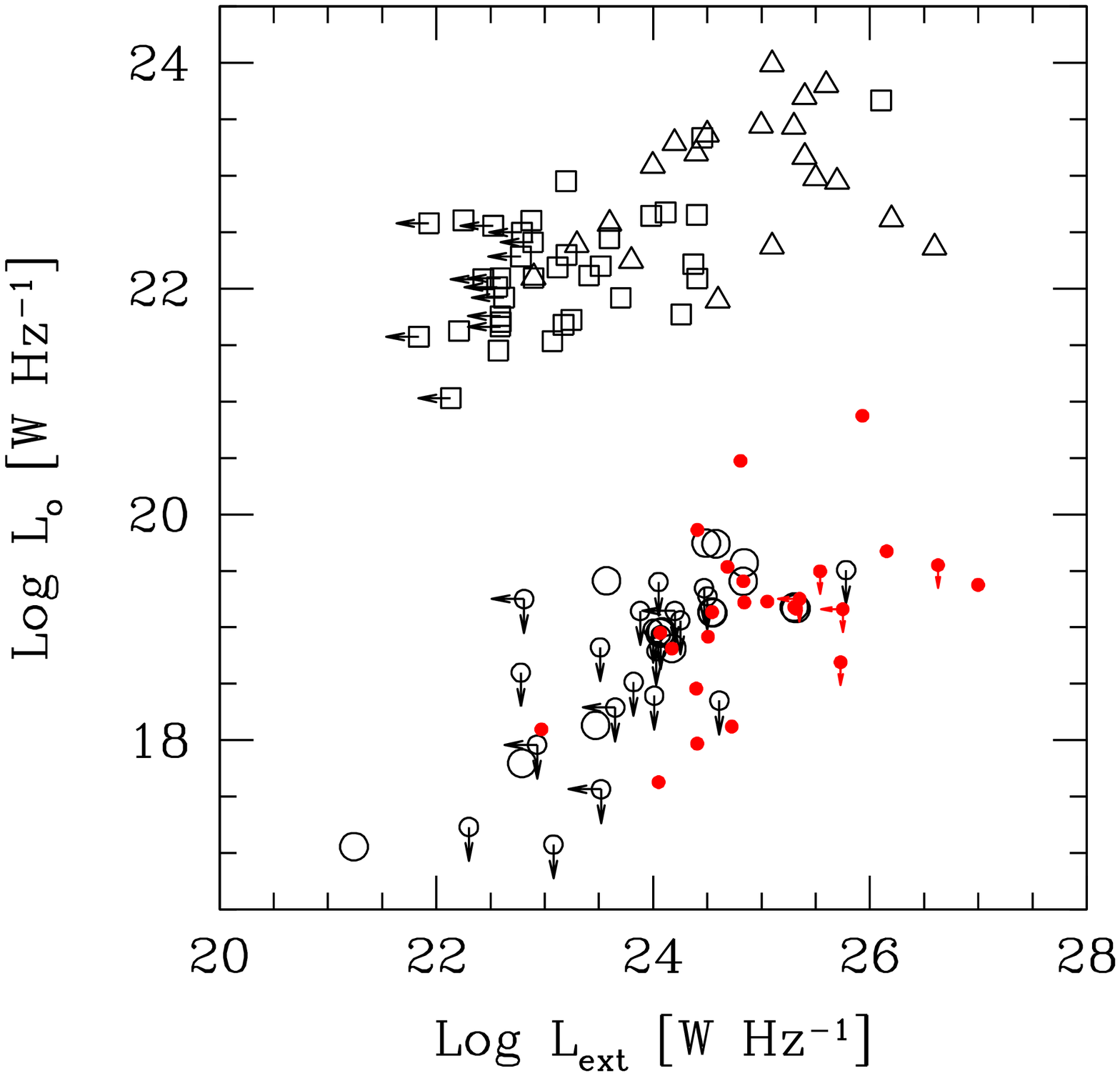,width=0.5\textwidth}
\psfig{figure=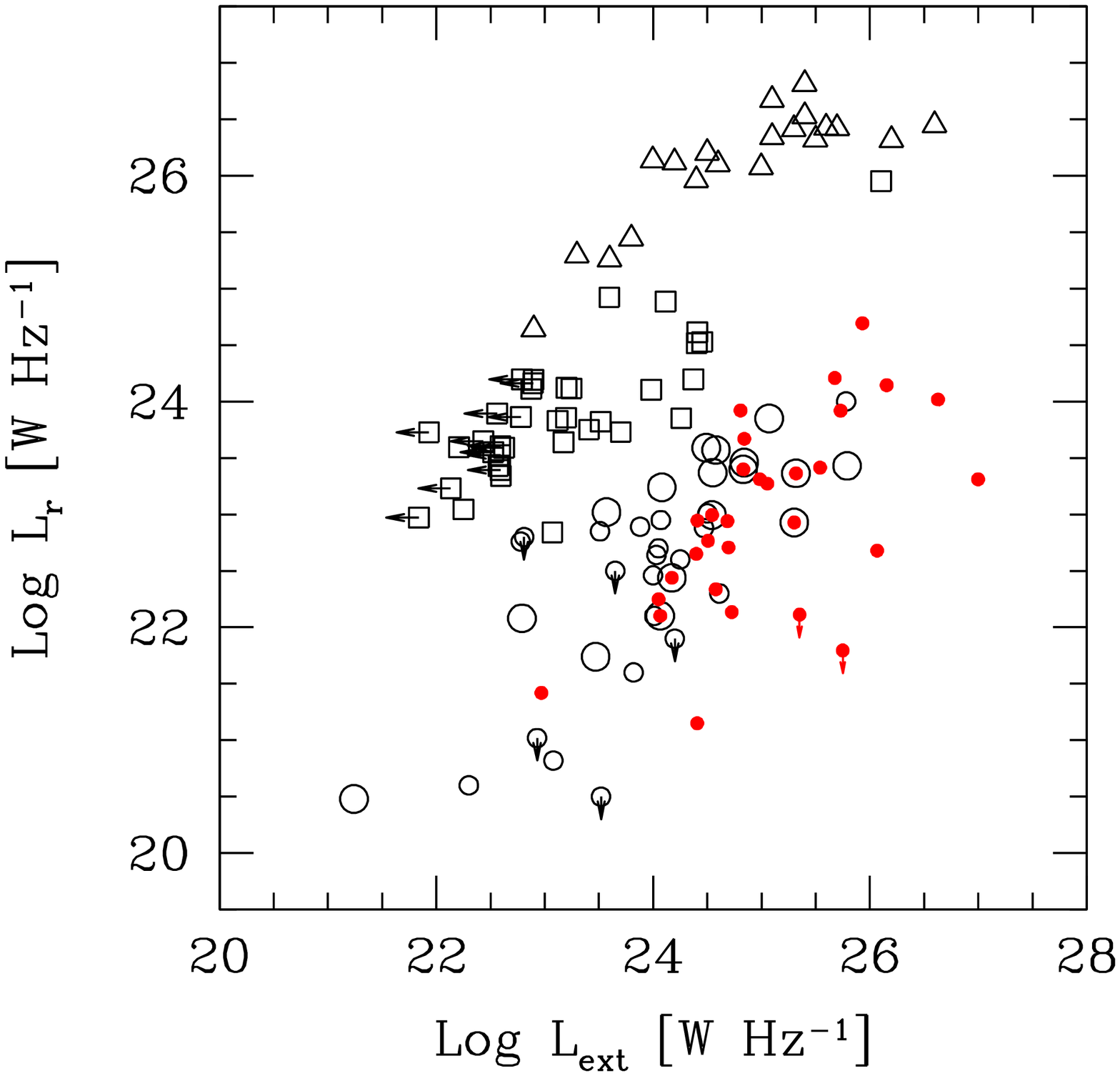,width=0.5\textwidth}}
\caption{Left panel: nuclear optical luminosity versus extended radio
power for radio galaxies and BL Lacs.  B2/FR~I sources are represented
by empty circles, while the filled circles are the 3C/FR~I sources
from Chiaberge et al. (\cite{chiaberge99}). Triangles and squares are
the 1 Jy and the {\sl Einstein} BL Lacs samples, respectively.  
Right panel: same
as the left panel for the core and extended radio luminosities.}
\label{ext}
\end{figure*}

In Fig.~\ref{ext} we plot the CCC optical and radio luminosity versus
the extended radio luminosities.  In this plane, the B2 sources
significantly improve the coverage at low extended luminosities with
respect to the 3C/FR~I sample studied by Chiaberge et
al. (\cite{chiaberge00b}).  Indeed, in the 3C/FR~I there is only one
object (3C272.1) with $L_{\rm ext} < 10^{24}$ W Hz$^{-1}$.  Conversely, in
the B2/HST sample, there are 21 galaxies below this value. In
particular the B2 sources cover the region of extended radio power
typical of HBL allowing a direct comparison over the whole range of
the FR~I/BL Lac populations.

To perform this comparison we consider BL Lacs selected from the 1 Jy catalogue
(Stickel et al. \cite{stickel91},
K\"uhr et al. \cite{kuhr}) and the {\it Einstein} Slew survey 
(Elvis et al.  \cite{elvis}, Perlman et
al. \cite{perl96}), which comprise 34 and 48 objects, respectively.
Of the 34
objects belonging to the 1 Jy sample, 32 are classified as LBL and 2
as HBL, while of the 48 X-ray selected BL Lacs, 40 are HBL and 8 are
LBL. The extended radio power is taken from Kollgaard et
al. (\cite{koll96}) and Perlman et al. (\cite{perl96}). 
On average HBL are characterized by lower extended radio luminosities
than LBL: the former sample extends from $\sim 10^{22}$ to 
$10^{24.5}$ W Hz$^{-1}$, the latter from 
$\sim 10^{23}$ W Hz$^{-1}$ to $10^{26}$ W Hz$^{-1}$.

In the $L_{\rm ext}/ L_{\rm o}$ plane FR~I nuclei and BL Lac lie in two well
separated stripes and the regions
covered by the two classes of BL Lacs are continuously connected. 
The optical luminosity of both BL Lacs and FR~I increases with the
extended power, with a similar logarithmic slope.
Let us consider the $L_{\rm ext}$ / $L_{\rm r}$ plane.
Radio galaxies follow a common trend, reflecting the well known
correlation between core and extended radio emission (Giovannini et
al. \cite{giovannini}). 
As expected, the two classes of BL Lacs
appear to have different core radio powers (likely due to selection
effects,  e.g. Giommi \& Padovani \cite{giopad94}), but the logarithmic 
slope of the BL Lac population as a whole is similar
to that measured in radiogalaxies.  
This is indeed what is expected if
the difference in core luminosities between BL Lacs and FR~I is due to
relativistic beaming with a bulk Lorentz factor independent of
$L_{\rm ext}$.  

It is clearly difficult to perform a detailed statistical comparison
between these two classes of objects as they are affected
by different selection biases. 
In particular, as discussed by Kollgaard et al. (\cite{koll96}),  
to adopt a survival analysis to account for the presence of upper limits,
we would have to use different quantities 
(extended luminosities for radio galaxies
and core luminosities for BL Lacs) as independent variables and the results
cannot be directly compared. 
We  preferred to follow their approach of evaluating 
the relationship between core and extended
luminosities of the two classes by determining 
the bisector from an ordinary least-squares procedure, treating upper limits
as detections. 
Note, however, that the physical parameter of interest,
the Doppler factor, that we estimate from this comparison 
depends only very weakly on the luminosity difference between
the two classes of objects and we thus believe that the 
uncertainties of this procedure do not affect significantly
our conclusions. 

The resulting luminosity differences between FR~I and BL Lac
are $L_{\rm BL}/L_{\rm FR~I}=10^{2.5}$ 
and $L_{\rm BL}/L_{\rm FR~I}=10^{3.9}$ in
the radio and optical band respectively.
We derive an average ratio of the Doppler factors between
the two classes of $\delta_{\rm BL}/\delta_{\rm FR~I}=
(L_{\rm BL}/L_{\rm FR~I})^{1/(p+\alpha)} \sim$ 18 (for a radio spectral index
$\alpha_{\rm r} = 0$, characteristic of self-absorbed sources, 
and $p=2$, as appropriate
if the emitting region is a continuous jet; for $p=3$, representing a 
moving sphere, $\delta_{\rm BL}/\delta_{\rm FR~I} \sim 7$).  Similarly, in
the optical band, $\delta_{\rm BL}/\delta_{\rm FR~I} \sim$ 20 
(for $\alpha_o =
1$ and $p=2$; for $p=3$,
$\delta_{\rm BL}/\delta_{\rm FR~I} \sim 9$). These values are 
consistent with those in the radio band, i.e. the different radio and
optical spectral indices account for the apparently different
amplification in the two bands.

In other words this comparison indicates that the differences in
luminosity between BL Lac and FR~I can be explained with a single
amplification factor over the whole range of extended luminosity in
both the radio and optical bands. This clearly provides further
support to the FR~I / BL Lac unifying scheme and to the interpretation
that in all cases we are seeing the synchrotron emission from a
relativistic jet.

Let us now consider how the difference in Doppler factors between BL
Lac and radio galaxies can be accounted for by a different viewing
angle of the same relativistically beamed emission region.  Assuming
the same intrinsic Lorentz factor and as average viewing angles
1/$\Gamma$ and 60$^{\circ}$ for BL Lac and FR~I, respectively, we
obtain $\delta_{\rm BL}/\delta_{\rm FR~I} = \Gamma^2/2$ and therefore $\Gamma
\sim 6$ for p=2( for p=3 $\Gamma \sim 4$).  As already noted by
CCCG00, this is significantly smaller
than what inferred from observational and theoretical considerations
on BL Lacs, which require $\Gamma \sim 15 - 20$ from both pair
opacity and broad band energy distribution constraints (e.g.  Dondi \&
Ghisellini \cite{dondi}, Ghisellini et al. \cite{gg98}, Tavecchio,
et al. \cite{taold}).  One possible interpretation of
such discrepancy, as suggested by CCCG00, is that a velocity structure
is present in the jet, with a fast spine (which dominates the emission
in BL Lac) surrounded by a slower layer (dominant in FR~I). 
 CCC99
estimated the typical value of $\Gamma$ for the layer in the simplest
(arbitrary) scenario where the two components have the same intrinsic
luminosity and spectra and derived $\Gamma_{\rm layer} \sim 1.5 - 2$.
Laing et al. \cite{laing99} from the radio core dominance in the B2 sample
of radio galaxies estimate $\Gamma_{\rm layer} \sim  2$ 
and $\Gamma_{\rm spine} \ge 9$,
close to the estimate of CCC99 but with  luminosity ratio 
$\frac{\rm I_s}{\rm I_l} \sim
6$. These figures however are still poorly constrained by the available data.

\section{Summary and conclusions}
\label{summary}

We discuss the optical properties of the nuclei of low luminosity
radio galaxies as derived from the snapshot HST images of the B2
sample.  A nuclear optical component is found in 18 out of the 57
observed galaxies and, in agreement with the results obtained from the
brighter 3C/FR~I sources, we found a correlation between fluxes (and
luminosities) of the optical and radio cores.  This provides further
support to the interpretation of a synchrotron origin of the optical
nuclear emission.  In the sources in which we failed to detect an
optical core, the optical limits are consistent with the
interpretation that their nuclei follow the same radio/optical
correlation, but their nuclear flux density is insufficient to be seen against
the galaxy background. However, the large fraction of undetected CCC does
not allow us to strengthen our conclusions on the lack of obscuring nuclear
matter (as it was in the case of the more powerful 3CR FR~I).

The radio/optical nuclear correlation for all FR~I extends down to an
optical luminosity of 10$^{18}$ W Hz$^{-1}$. This value can be adopted
as an upper limit to any emission from the AGN not directly related to
their relativistic jets, such as e.g. from the accretion disk.  For a
10$^9$ M\sun\ central black hole, this corresponds to a fraction $\sim
3\times 10^{-8}$ of the Eddington luminosity in the optical band and
it provides a clear evidence that accretion in these low luminosity
radio galaxies occurs at a very low rate and/or radiation efficiency.

In the framework of the FR~I/BL Lacs unified scheme, the direct
comparison of the optical nuclear properties of radio galaxies with
their putative aligned counterparts provides a quantitative method of
testing this unification model.  
The inclusion of the B2 sources significantly improves the coverage
towards low extended luminosities with respect to the 3C/FR~I sample
and thus it is now possible to compare the properties of the two
populations over their whole range of radio power.

This comparison indicates that the differences in luminosity between
BL Lac and FR~I can be explained with a single amplification factor
over the whole range of extended radio luminosity in both the radio
and optical bands, again supporting the interpretation that in both
cases we are seeing relativistic jet synchrotron emission.  The
corresponding bulk Lorentz factor results (for typical viewing angles)
significantly smaller than derived from spectral energy properties.
This support the interpretation that a velocity structure is present
in the jet, with a fast spine surrounded by a slower layer where the
layer is responsible for the bulk of the intrinsic emission. 

\begin{acknowledgements}
The authors acknowledge the Italian MURST for financial support.  This
research has made use of the NASA/IPAC Extragalactic Database (NED)
which is operated by the Jet Propulsion Laboratory, California
Institute of Technology, under contract with the National Aeronautics
and Space Administration.
\end{acknowledgements}

\end{document}